\documentclass[%
 aip,
 amsmath,amssymb,
 reprint,%
]{revtex4-2}
\usepackage{float}
\usepackage{graphicx}
\usepackage{dcolumn}
\usepackage{bm}

\usepackage[T1]{fontenc}
\usepackage[english]{babel}
\usepackage{mathptmx}
\usepackage{xcolor}

\newcommand{\be}{\begin{equation}}
\newcommand{\ee}{\end{equation}}
\newcommand{\bea}{\begin{eqnarray}}
\newcommand{\eea}{\end{eqnarray}}

\begin{document}


\title[]{Insights into elastic properties of coarse-grained DNA models: q-stiffness 
of cgDNA vs. cgDNA+}

\author{Wout Laeremans}
\affiliation{Soft Matter and Biological Physics, Department of Applied Physics, and Institute for Complex Molecular Systems,
Eindhoven University of Technology, P.O. Box 513, 5600 MB Eindhoven, Netherlands}
\affiliation{Soft Matter and Biophysics Unit, KU Leuven, Celestijnenlaan
200D, 3001 Leuven, Belgium}
\affiliation{UHasselt, Faculty of Sciences, Data Science Institute, Theory Lab, Agoralaan, 3590 Diepenbeek, Belgium}
\author{Midas Segers}
\author{Aderik Voorspoels}
\affiliation{Soft Matter and Biophysics Unit, KU Leuven, Celestijnenlaan
200D, 3001 Leuven, Belgium}
\author{Enrico Carlon}
\email{enrico.carlon@kuleuven.be}
\affiliation{Soft Matter and Biophysics Unit, KU Leuven, Celestijnenlaan
200D, 3001 Leuven, Belgium}
\author{Jef Hooyberghs}
\email{jef.hooyberghs@uhasselt.be}
\affiliation{UHasselt, Faculty of Sciences, Data Science Institute, Theory Lab, Agoralaan, 3590 Diepenbeek, Belgium}

\date{\today}

\begin{abstract}
Coarse-grained models have emerged as valuable tools to simulate long DNA 
molecules while maintaining computational efficiency. These models aim at 
preserving interactions among coarse-grained variables in a manner that
mirrors the underlying atomistic description. We explore here a method for 
testing coarse-grained vs. all-atom models using 
stiffness matrices in Fourier space ($q$-stiffnesses), which are particularly 
suited to probe DNA elasticity at different length scales. 
We focus on a class of coarse-grained rigid base DNA models known as cgDNA and 
its most recent version cgDNA+. Our analysis shows that while cgDNA+ follows closely 
the $q$-stiffnesses of the all-atom model, the original cgDNA shows some deviations 
for twist and bending variables which are rather strong in the $q \to 0$ (long length
scale) limit. The consequence is that while both cgDNA and cgDNA+ give a suitable
description of local elastic behavior, the former misses some effects which manifest
themselves at longer length scales. In particular, cgDNA performs poorly on the twist 
stiffness with a value much lower than expected for long DNA molecules. Conversely, 
the all-atom and cgDNA+ twist is strongly length scale dependent:
DNA is torsionally soft at a few base pair distances, but becomes more rigid at 
distances of a few dozens base pairs. Our analysis shows that the bending persistence 
length in all-atom and cgDNA+ is somewhat overestimated.
\end{abstract}

\maketitle

\section{Introduction}

All-atom simulations have played a fundamental role in our understanding of DNA 
mechanics. Despite remarkable advances in computing power, such simulations 
are still limited to sequences of few tens of base pairs (bp) and to times 
spanning about $1~\mu\rm{s}$ \cite{pasi14}. Sometimes advanced sampling techniques 
can be used to alleviate some of these limitations by simulating rare 
events and extreme conformations \cite{curu09,spir12,karo14,pegu17,voor23}. 
However, the large computational cost remains a heavy burden.
When atomistic details are not of crucial importance, coarse-grained models 
\cite{ould10,henr18,dans10,fosa16,chak18,asse22} represent a valid alternative 
to all-atom simulations as they can handle $100-1000$~bp long molecules for 
considerably longer time scales. In these models, lower-resolution, 
coarse-grained beads interact with each other through potentials that are 
parametrized to reproduce the thermodynamic, structural, and mechanical 
properties of DNA. Several coarse-grained DNA models neglect sequence dependent 
effects, but can be used to explore conformational changes, strand 
dissociation/hybridization, supercoils formation and other 
effects \cite{fred14,mate15,cord17,coro18,cara19}.

\begin{figure}[t]
\includegraphics[width=0.8\linewidth]{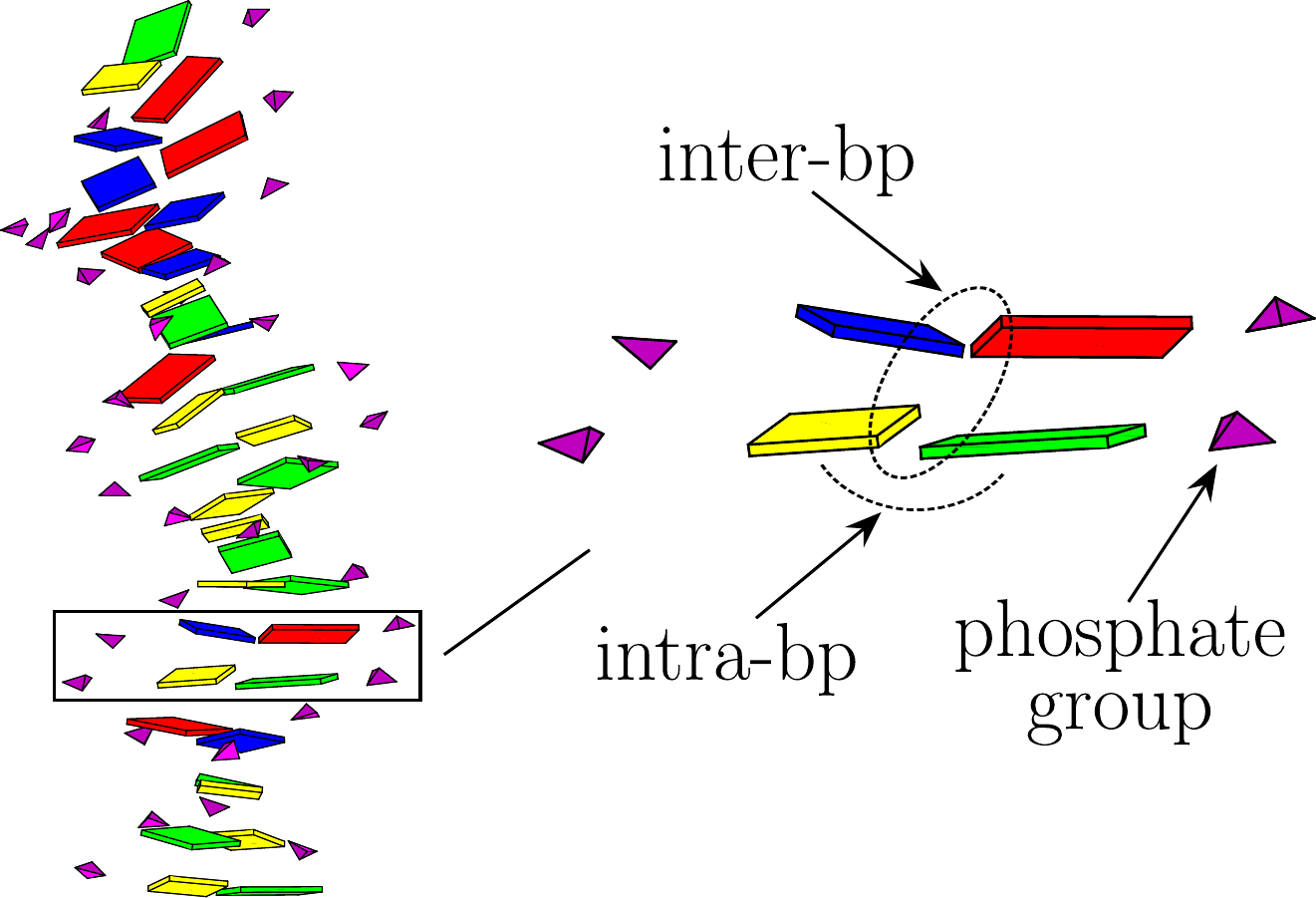}
\caption{Rigid base models (as cgDNA \cite{petk14}) represent bases in
each B-DNA molecule as rigid objects, here plotted as rectangles with
the following color code: A (red), T (blue), C(yellow) and G(green).
In the original cgDNA-model, a DNA conformation is described by $12$
coordinates, $6$ of which are intra-bp coordinates and $6$ inter-bp
coordinates, parametrizing rotations and translations about the rigid
units.  The image shows a rigid base model conformation sampled from
cgDNA+ \cite{pate19,shar23}, which takes explicitly into account phosphate
groups, here shown as purple pyramids.}
\label{fig:snapshot-cgDNA} 
\end{figure}

A very good account of sequence-dependent effects is given
by coarse-grained rigid base DNA models \cite{petk14,shar23}.
Such models describe bases as rigid bodies and a DNA conformation
by means of rotations and translation between the rigid units, see
Fig.~\ref{fig:snapshot-cgDNA}.  The rigid base models describe DNA
molecules in their canonical B-form and at a fixed temperature, the base
pairs do not dissociate and self-avoidance effects are not included. The
interactions between the rigid bases are usually encoded by harmonic
potentials \cite{petk14,shar23} and can be represented by a stiffness
matrix. Recently also multimodal interactions were proposed \cite{walt20}.
The standard cgDNA model \cite{petk14} parametrizes the DNA conformations
using the canonical twelve coordinates, defined by the Tsukuba convention
\cite{olso01}.  Six of these are intra base pair coordinates (buckle,
propeller, opening, shear, stretch, stagger) and six are inter base
pair (or junction) coordinates (tilt, roll, twist, shift, slide, rise).
In the more recent cgDNA+ \cite{pate19,shar23}  the phosphate groups are
treated explicitly, which brings the number of coordinates to $24$ per
bp, see Fig.~\ref{fig:snapshot-cgDNA}. The parameters of the cgDNA/cgDNA+
models were tuned to all-atom simulations that are used as references. For
this purpose, the model stiffness matrix was determined by minimizing
the distance between the associated covariance matrix of the model and
that of the all-atom simulations \cite{gonz13}.

The focus of this paper is on mechanical properties of DNA that involve 
couplings beyond nearest-neighbor sites, following recent work along this 
line \cite{skor21,sege22}. For this purpose we study the stiffness 
of cgDNA and cgDNA+ in Fourier space 
($q$-stiffness) and compare it with all-atom simulations. Such quantities 
provide considerable insights about the elastic properties of the molecule 
at different length scales \cite{skor21}. We show that some degrees of freedom 
are stiffer/softer at the base pair level, as compared to the long wavelength 
limit ($q \to 0$), describing the asymptotic stiffness of very long molecules.
Our analysis shows that cgDNA+ considerably improves over the earlier version 
cgDNA. The cgDNA+ model shows a remarkable quantitative agreement with all-atom 
data for all $q$-stiffnesses of the twelve inter- and intra-bp coordinates.
Besides the specific analysis of rigid base models, this work provides a 
methodological example for testing coarse-grained DNA models via $q$-stiffnesses.
This paper is organized as follows. Section~\ref{sec:RBM} reviews the main
properties of the rigid base model and introduces the stiffnesses in Fourier
space, also referred to as $q$-stiffnesses, which are computed and discussed
in Sec.~\ref{sec:q-stiff}. Section~\ref{sec:length-dep} links the $q$-stiffnesses
to length scale dependent elasticity, focusing on torsional and bending
persistence lengths. Section~\ref{seq:correlations} focuses on the real-space 
two-point correlation function of several rigid base parameters and their 
associated $q$-stiffness. Section~\ref{sec:discussion} discusses the results
obtained and presents some conclusions.

\section{Rigid base models}
\label{sec:RBM}

We consider here a description based on the canonical rigid base coordinates
introducing a twelve dimensional vector $\Delta_n$, with $n=0,1,\ldots N-1$ 
labeling the sites along the sequence. The elements of the vector $\Delta_n$ 
represent the deviation of the twelve rigid base coordinates with respect to 
their average values, such that $\langle \Delta_n \rangle = 0$, with 
$\langle . \rangle$ denoting thermal averaging.  
Small deformations from equilibrium are usually described within the
harmonic approximation \cite{dohn21}, with an energy given by
\begin{equation}
\beta E = \frac{a}{2} \sum_n \sum_m
{\Delta}^\intercal_n M^{(n)}_m {\Delta}_{n+m},
\label{eq:energy_real}
\end{equation}
with $\beta=1/k_BT$ the inverse temperature, $k_B$ the Boltzmann constant and 
$a$ the average distance between consecutive base pairs ($a=0.34$~nm for DNA). 
$M^{(n)}_m$ are $12 \times 12$ stiffness matrices and depend on the sequence
composition, reflected in the dependence on the site index $n$. The above 
model allows for distal couplings between variables at distinct sites $n$
and $n+m$.

As our primary interest will be to highlight the nature of the distal
couplings, we will neglect sequence dependent effects and ignore the $n$
dependence in $M_m^{(n)}$. Assuming an effective translationally invariant 
DNA model (obtained by averaging over different sequences) it is convenient 
to introduce a discrete Fourier transform of the canonical rigid base 
coordinates as follows
\begin{equation}
    \widetilde \Delta_q = \sum_{n=0}^{N-1}  \Delta_n\, e^{-2\pi i q n/N},
    \label{def:Deltaq}
\end{equation}
where $q = -(N-1)/2, -(N-3)/2, \ldots ,(N-3)/2, (N-1)/2$ (for $N$ odd). 
As the vector $\Delta_n$ is real, complex conjugation gives
$\widetilde \Delta_q^* = \widetilde \Delta_{-q}$.
The model \eqref{eq:energy_real} in $q$-space then takes the form 
\begin{equation}
\beta E = \frac{a}{2N} \sum_q 
{\widetilde\Delta}^\dagger_{q} \widetilde{M}_q {\widetilde\Delta}_q,
\label{eq:energy_fourier}
\end{equation}
where $\dagger$ denotes Hermitean conjugation and where the matrix
$\widetilde{M}_q$ is obtained by taking the Fourier transform of $M_m$ \cite{skor21}
\begin{equation}
    \widetilde{M}_q  = \sum_m  M_m \, e^{-2\pi i q m/N}.
    \label{eq:mtilde}
\end{equation}
The assumed translational invariance for a finite system only applies with periodic
boundary conditions. The open DNA molecule will also have additional boundary
terms, which do not influence the large $N$ bulk behavior.
The matrix $\widetilde{M}_q$ is $12 \times 12$ and, for convenience, we organize 
it in $16$ submatrices as follows
\begin{align}
\widetilde{M}_q &\equiv 
\begin{bmatrix}
\widetilde{A}_q & \widetilde{B}_q & \widetilde{C}_q & \widetilde{D}_q \\
\widetilde{E}_q & \widetilde{F}_q & \widetilde{G}_q & \widetilde{H}_q \\
\widetilde{I}_q & \widetilde{J}_q & \widetilde{K}_q & \widetilde{L}_q \\
\widetilde{N}_q & \widetilde{O}_q & \widetilde{P}_q & \widetilde{Q}_q
\end{bmatrix}.
\label{eq:Mq}
\end{align}
where $\widetilde{A}_q$, $\widetilde{B}_q$ \ldots $\widetilde{Q}_q$ are 
$3 \times 3$ matrices. The order of the 12 coordinates is chosen to be: 
tilt, roll, twist, shift, slide, rise, buckle, propellor, opening, shear, 
stretch, stagger. For example, the first element on the diagonal of 
$\widetilde{F}_q$ is the $q$-stiffness of the ``shift'' variable. 
The elements $1$ to $6$ and $7$ to $12$ correspond to inter- and intra-bp 
coordinates, respectively. Off-diagonal elements are cross-coupling terms 
between different coordinates as e.g. ``shift-slide'' or ``roll-twist''. 
The stiffness matrix can be obtained from the inversion of the covariance 
matrix \cite{skor21}
\begin{equation}
\widetilde{M}_q = \frac{N}{a}
\left\langle\widetilde{\Delta}_q\widetilde{\Delta}_q^\dagger\right\rangle^{-1},
    \label{eq:covariance}
\end{equation}
where, in our case, $\widetilde{\Delta}_q$ is obtained from simulations. 
$\widetilde{\Delta}_q$ and $\widetilde{\Delta}_q^\dagger$ are $12$-dimensional 
column and row vectors. The product in \eqref{eq:covariance} gives a $12 \times 12$ 
matrix. The Monte Carlo (cgDNA/cgDNA+) sampling or the molecular dynamics (MD) simulations 
(all-atom) produces real space configurations expressed in terms of the $12$ canonical
coordinates. For every simulated sample, this gives a vector $\Delta_n$ for each site 
$n = 0,1,...,N-1$ of the DNA sequence. These are Fourier transformed to give 
$\widetilde{\Delta}_q$. Using Eq.~\eqref{eq:covariance}, the 
matrix $\widetilde{M}_q$ is deduced. In general, the elements of $\widetilde{M}_q$
have real and imaginary parts. From Eq.~\eqref{eq:mtilde}, it follows that the real 
part of the elements of $q$ are symmetric in $q$, while the imaginary parts are 
antisymmetric.

\begin{figure}[t]
    \centering
    \includegraphics[width=\linewidth]{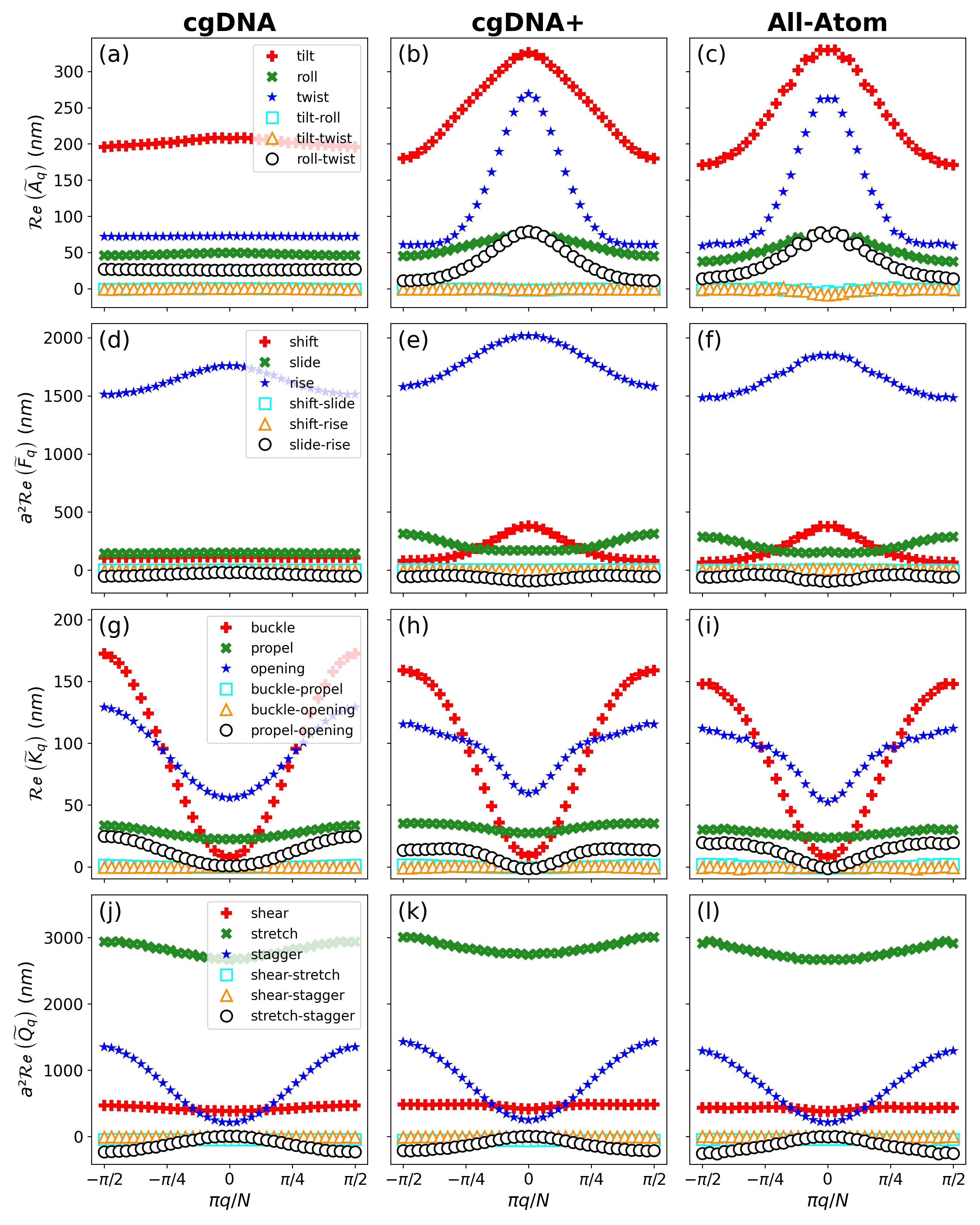}
\caption{Real part of $q$-stiffnesses of inter- and intra-basepair degrees
of freedom obtained from Monte Carlo sampling the cgDNA and cgDNA+
models and all-atom MD simulations (left, center and right columns,
respectively). The sub-blocks $\widetilde{A}_q$, $\widetilde{F}_q$,
$\widetilde{K}_q$ and $\widetilde{Q}_q$ of $\widetilde{M}_q$ are
shown \eqref{eq:Mq}. Stiffnesses of translational degrees of freedom
($\widetilde{F}_q$, $\widetilde{Q}_q$) are rescaled by a factor $a^2$
($a=0.34$ nm) in order to express stiffnesses in units nm in all plots.
cgDNA+ stiffnesses are in very good agreement with MD simulations.
The largest differences between cgDNA and cgDNA+ are for tilt and twist
coordinates (block $\widetilde{A}_q$, (a,b,c)). cgDNA lacks the peak at
$q=0$ for tilt, twist and twist-roll couplings, which is present in cgDNA+
and all-atom data.}
\label{fig:q-stiff} 
\end{figure}


\section{$q$-stiffness in cgDNA(+) vs. MD simulations}
\label{sec:q-stiff}

Figure~\ref{fig:q-stiff} shows a plot of the real parts of the
elements of $\widetilde{A}_q$, $\widetilde{F}_q$, $\widetilde{K}_q$ and
$\widetilde{Q}_q$, the four $3 \times 3$ diagonal sub-block matrices of
$\widetilde{M}_q$ which was obtained from Eq.~\eqref{eq:covariance}. The
three columns report the data from cgDNA, cgDNA+ and all-atom simulations
which are obtained from averaging over two different sequences, see
details in Appendix \ref{appendix:all_atom}.  The cgDNA/cgDNA+ data are
generated using a Monte Carlo sampling, while the all-atom data are from
averaging a $100~\text{ns}$ MD simulation (data from \cite{sege23}). There
is a remarkable agreement here between cgDNA+ and all-atom data,
and a considerable improvement from cgDNA to cgDNA+ for some, but not
for all, matrix elements. The strongest differences between cgDNA and
cgDNA+ is in the twist and bend (tilt/roll) coordinates, which are in
the block $\widetilde{A}_q$ and shown in Fig.~\ref{fig:q-stiff}(a,b,c).
For these coordinates the cgDNA stiffnesses are very weakly $q$-dependent,
while a strong dependence for tilt and twist is observed in the cgDNA+
and the all-atom data.

Large variations in stiffnesses for different $q$ indicate that the
elastic response of the molecule strongly depends on the length scale
at which it is probed \cite{esla16,skor21,fosa23}, as discussed in more
details in Sec.~\ref{sec:length-dep}. For tilt and twist, cgDNA fits well
the short scale behavior which corresponds to $|q|\approx N/2$, i.e. the
two edges of the graphs in Fig.~\ref{fig:q-stiff}. However, the cgDNA data
lacks the sharp peaks at $q=0$ found in cgDNA+ and all-atom simulations
(b,c). The $q=0$ corresponds to the long length scale behavior, as
discussed in Sec.~\ref{sec:length-dep}. The good agreement between the
$q$-stiffnesses of cgDNA and cgDNA+ at $|q| \approx N/2$ stems from
the fitting of local couplings of the all-atom data. To improve cgDNA,
i.e. to reproduce the $q$-dependence observed in all-atom data, one would
need to account for distal components by adding non-vanishing matrix
elements in $M_m$ in \eqref{eq:mtilde}. These effective distal couplings
are generated via the integration of phosphate coordinates in cgDNA+.

\begin{figure}[t]
    \centering
    \includegraphics[width=\linewidth]{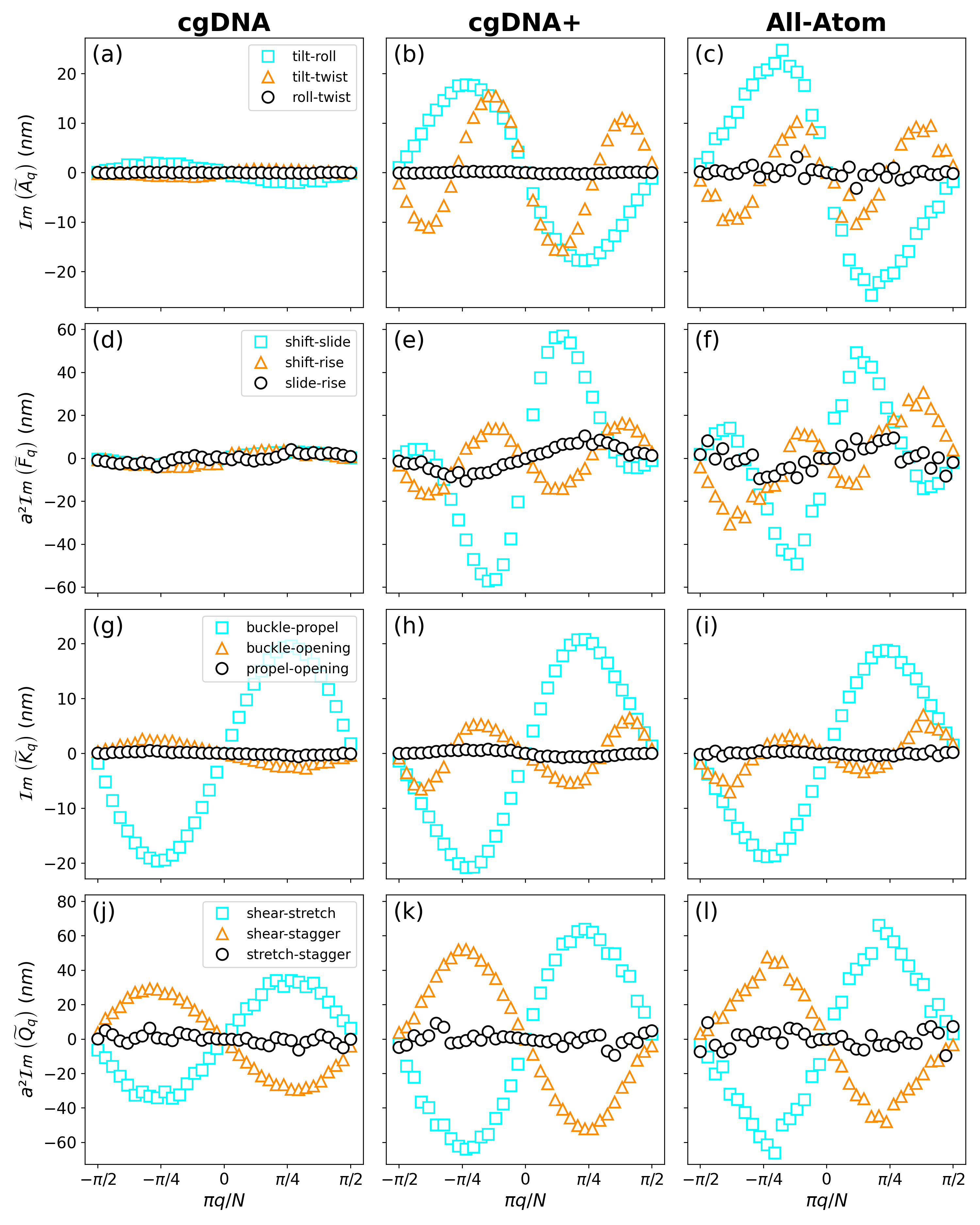}
\caption{Imaginary part of $q$-stiffnesses of rigid base coordinates
for cgDNA/cgDNA+ and all-atom simulations, showing the same sub-blocks of 
the stiffness matrix as in Fig.~\ref{fig:q-stiff}. Similar to the real 
counterparts, the imaginary parts of the cgDNA+ stiffnesses (middle column)
are in excellent agreement with all-atom MD simulations (right column), whereas the cgDNA-model does not reproduce all-atom data. The lack of 
structure is most notably for the inter-basepair coordinates (a-f).}
\label{fig:q-stiff-imag} 
\end{figure}

Figure~\ref{fig:q-stiff}(d,e,f) compare the translational inter-bp
stiffnesses. While there is good agreement in all three models
for the rise stiffness (the stiffer of the translational inter-bp
coordinate), the cgDNA+ and all-atom data show remarkable similarity
in the $q$-dependence for slide and shift. Stiffnesses of intra-bp
coordinates (Fig.~\ref{fig:q-stiff}(g-l)) for all three models are
in good agreement with each other, which indicates that the additional
phosphate degrees of freedom of cgDNA+ have a weak effect on the intra-bp
coupling. Nevertheless, some minor differences in shape can be observed,
most visible in case of the opening stiffness. We note that there are
two different types of $q$-dependences in Fig.~\ref{fig:q-stiff}. Some
stiffnesses have a maximum at $q=0$ (inter-bp couplings (a-f)), while
others have a minimum at $q=0$ (intra-bp couplings (g-l)). A maximum
at $q=0$ indicates that the corresponding coordinate is soft at short
scale and becomes stiffer at longer scales, while the opposite is true
for a minimum at $q=0$.

Figure~\ref{fig:q-stiff-imag} shows a plot of the imaginary parts of
the blocks $\widetilde{A}_q$, $\widetilde{F}_q$, $\widetilde{K}_q$ and
$\widetilde{Q}_q$, of the matrix $\widetilde{M}_q$. Symmetry imposes
vanishing imaginary part to all diagonal elements, hence only the three
off-diagonal components of each block matrix are shown in the plots of
Fig.~\ref{fig:q-stiff-imag}.  Similarly to their real counterparts in
Fig.~\ref{fig:q-stiff}, we note a much weaker $q$-dependence for cgDNA as
compared to cgDNA+ and all-atom data.  The weak $q$-dependence in cgDNA is
again very striking for the elements of the blocks $\widetilde{A}_q$ and
$\widetilde{F}_q$, depicting the stiffnesses of the inter-bp coordinates
(see plots (a-f)). Again cgDNA+ and all-atom data appear to be in
excellent agreement with each other.

\section{Length-scale dependent elasticity}
\label{sec:length-dep}

A model characterized by a $q$-dependent stiffness has elastic properties
which depend on the length scale at which they are probed\cite{skor21}. In
this section we elaborate on the twist and bending properties. To
introduce the concept we evaluate the response of the system to a pure
twist perturbation. We start with the partition function $Z$ and focus
on the excess twist, which we denote with $\widetilde{\Omega}_q$, by
integrating out all other eleven variables in the rigid base model. This
gives
\begin{equation}
    Z = \int \prod_q d \widetilde{\Delta}_q \,\, e^{-\frac{a}{2N} \sum_q 
    {\widetilde\Delta}^\dagger_{q} \widetilde{M}_q {\widetilde\Delta}_q}
    = \int \prod_q d \widetilde{\Omega}_q \,\, e^{-\beta E_\text{twist}}.
\end{equation}
As the full model is Gaussian, the integration over all degrees of
freedom but twist gives an effective one dimensional model for twist
elasticity which is still Gaussian. The energy is thus quadratic in
$\widetilde{\Omega}_q$ and takes the form
\begin{equation}
\beta E_{\text{twist}} = \frac{a}{2N} 
\sum_q \widetilde{\cal C}_q  \left| \widetilde{\Omega}_q \right|^2,  
\label{eq:en-twist}
\end{equation}
where the scalar $\widetilde{\cal C}_q$ is the stiffness of the mode $q$.
We note that in general $\widetilde{\cal C}_q \neq (\widetilde{M}_q
)_{33}$ because of off-diagonal couplings between twist and other
coordinates.  More precisely, $\widetilde{\cal C}_q$ can be obtained as a
Schur complement of the $12 \times 12$ stiffness matrix $\widetilde{M}_q$
\cite{pate19,fosa23}.

A physical way to probe the length-scale dependent twist response is to
apply a torque $\tau$ to a sub-sequence of $m$ base pairs. The energy
of the system gets an extra term and becomes, in real space coordinates,
\begin{equation}
    \beta E^{\tau} = \beta E_{\text{twist}} - 
      \beta \tau a \sum_{n=1}^m \Omega_n,
\label{eq:torque}
\end{equation}
with $a \sum_{n=1}^m \Omega_n$ the total excess twist angle of the DNA
segment of $m$ base pairs.  To calculate the partition function $Z$ for
\eqref{eq:torque} it is convenient to switch to Fourier space coordinates
$\widetilde{\Omega}_q$.  presented in previous work \cite{skor21}. The
total free energy $F=-k_B T \log Z$ is then given by
\begin{equation}
    F(\tau) = F_0 - \frac{m \beta \tau^2}{2 {\cal{C}}_m},
\label{eq:free_en}
\end{equation}
where ${\cal{C}}_m$ is introduced as the variable of interest: the
torsional stiffness related to a torque applied to $m$ subsequent
base pairs. Note that the free energy \eqref{eq:free_en}, as an
extensive quantity, contains a torque response term proportional to $m$.
The crucial point here is that in general ${\cal{C}}_m$ depends on $m$:
the torsional stiffness depends on the length scale $m$ at which the
twist elasticity is probed.  The torsional stiffness is given by (the
details of the derivation can be found in
\cite{skor21})
\begin{eqnarray}
    \frac{1}{{\cal{C}}_m} &=& 
    \frac{a}{m\pi} \int_{-\pi/2}^{+\pi/2} 
    \frac{\sin^2 my}{\sin^2 y} \, \, 
    \frac{\langle | \widetilde{\Omega}_{q} 
    |^2  \rangle}{N}
    \, dy
    \nonumber \\
    &=&
    \frac{1}{m\pi} \int_{-\pi/2}^{+\pi/2} 
    \frac{\sin^2 my}{\sin^2 y} \, \, \frac{dy}{\widetilde{\cal C}_{q}},
\label{eq:integral}
\end{eqnarray}
where we introduced the rescaled momentum $y=\pi q/N$ and replaced the
discrete sum in $q$ with a continuous integral, assuming $N \to \infty$.
The asymptotic behavior $m \gg 1$ of \eqref{eq:integral} has been
discussed in prior work \cite{sege21} and it is given by
\begin{equation}
\label{eq: 1/m behaviour of C_m }
    {{\cal{C}}_m} = \frac{m}{m + A} \, \widetilde{\cal C}_{q=0}
    + {{O}}\left( e^{-m/\lambda}\right), 
\end{equation}
with $A$ a scale factor. Apart from an exponentially small correction ,
${\cal{C}}_m$ has a leading algebraic $1/m$ decay to $\widetilde{\cal
C}_{q=0}$.  At long length scales ($m \to \infty$) the integrand in
\eqref{eq:integral} is increasingly peaked around $y=0$, hence the
torsional stiffness of an infinitely long chain is that of the mode $q=0$.

\begin{figure}[t]       
\includegraphics[width=\linewidth]{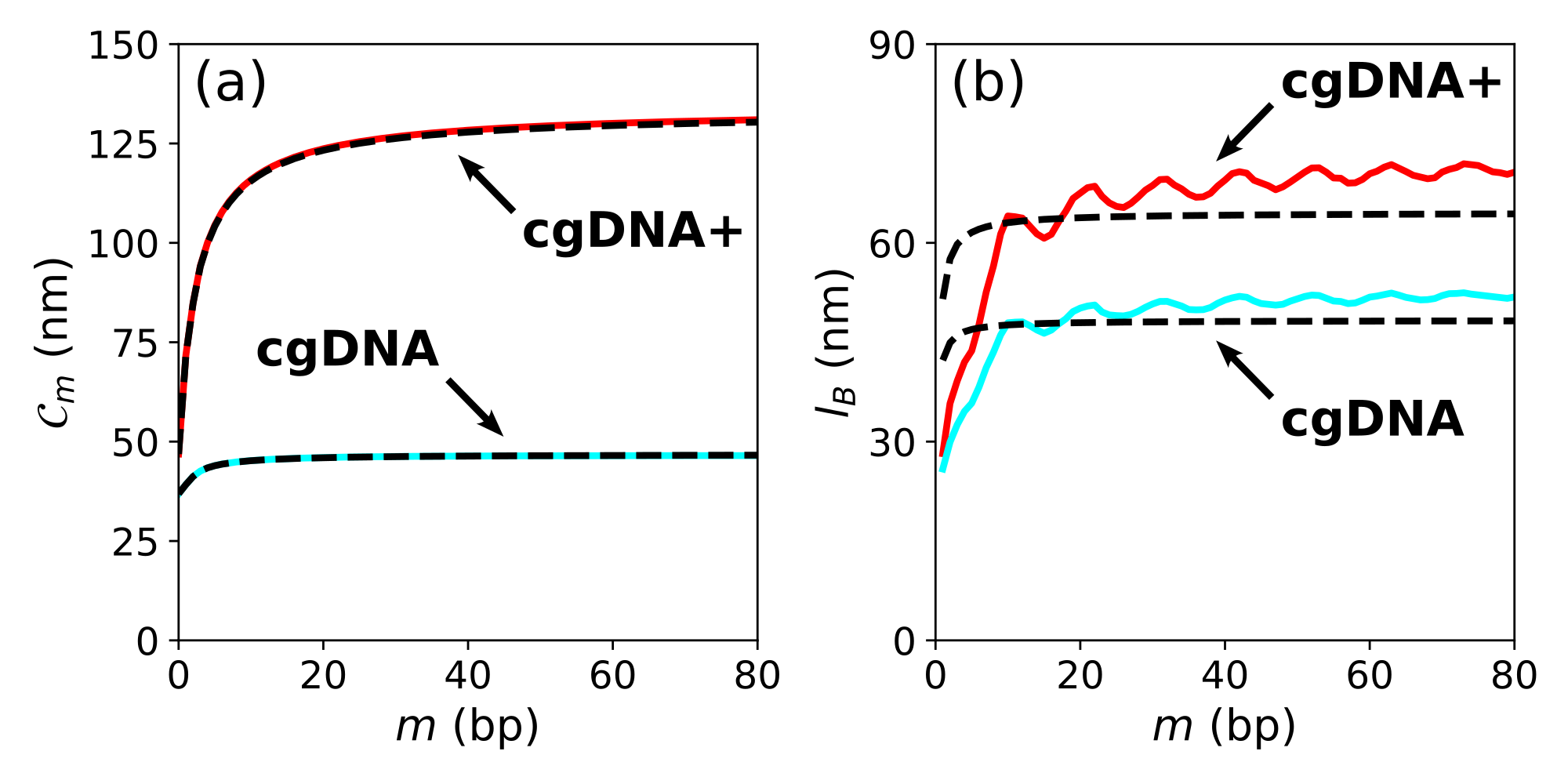}
\caption{(a) Solid lines: Torsional stiffness for cgDNA and cgDNA+,
obtained from the twist correlation function $\langle \cos\left( \sum_{i
= n}^{n+m-1} a \Omega_i \right) \rangle = e^{-am/(2{{\cal{C}}_m})}$
averaging over $100$ random sequences with length $500$~bp. Dashed lines:
Data obtained from the numerical integration of \eqref{eq:integral}. (b)
Solid lines: Bending persistence length calculated for cgDNA and cgDNA+
from the tangent-tangent correlation function $\langle \hat{t}_n \cdot
\hat{t}_{n+m} \rangle = e^{-am/l_B(m)}$, with $\hat{t}_n$ the unit tangent
vector at site $n$. Dashed lines: Bending persistence length $l_B$ as
obtained from the numerical integration of \eqref{eq:integral_lB}. We
notice a difference between the two estimates, whose origin is discussed
in the text. The extrapolated bending persistence length of cgDNA+ is
above the consensus value $l_B \approx 50~nm$, while the cgDNA value
is rather consistent with experiments. As in (a) the DNA appears to be
stiffer at longer scales.}
\label{fig:C-stiff} 
\end{figure}

With this concept in place, we analyzed the torsional
stiffness ${{\cal{C}}_m}$ of the cgDNA and cgDNA+
models. Figure~\ref{fig:C-stiff}(a) summarizes the results.
The dashed lines show the stiffness obtained from a numerical
integration of \eqref{eq:integral}. As a comparison, the solid lines
were obtained from real space data, not by applying a torque, but from
the correlation function \cite{skor21} $\langle \cos\left( \sum_{i =
n}^{n+m-1} a \Omega_i \right) \rangle = e^{-am/(2{{\cal{C}}_m})}$,
averaging over $100$ independent sequences of length $500$~bp. Both
the cgDNA and cgDNA+ model show an explicit $m$-dependence, they have
similar values of the stiffnesses at the shortest length scale ($m=1$).
For high $m$-values the curves obey equation (\ref{eq: 1/m behaviour
of C_m }) and decay to an asymptotic ${\cal{C}}$-value listed in
Table~\ref{tab:persistence_lengths}. cgDNA+ shows a strong length-scale
dependence reaching asymptotically a value which is close to (about $20\%$
higher) the value ${\cal C} = 110$~nm measured in single molecule magnetic
tweezers (MT) experiments \cite{brya03,lipf10,gao21}.  These experimental
values can be considered as the high-$m$ limit since such devices measure
the torsional response of a several Kbp long molecule, using a magnetic
bead attached to one of its ends, while its other end is attached to a
solid surface. cgDNA instead, shows a much weaker length dependence of
the torsional stiffness, with an asymptotic value ${{\cal{C}}_{m \to
\infty}} = 47~\text{nm}$, which is well-below the experimental results.

We now discuss length-scale dependence of bending deformations.
In Ref.~\onlinecite{skor21} the following expression for the bending 
persistence length $l_B$ was derived 
\begin{eqnarray}
    \frac{1}{l_B} = \frac{a}{m\pi} \int_{-\pi/2}^{+\pi/2} 
    \frac{\sin^2 my}{\sin^2 y} \, \, 
    \frac{\Psi_{q+\Delta q} + \Psi_{q-\Delta q}}{N} \, dy,
    \label{eq:integral_lB}
\end{eqnarray}
in which again $y=\pi q/N$ and 
\begin{eqnarray}
    \Psi_q = \frac{1-\cos(a \omega_0)}{2(a \omega_0)^2}
    \langle
    |\widetilde{\tau}_q|^2 +
    |\widetilde{\rho}_q|^2
    \rangle,
\end{eqnarray}
where $\widetilde{\tau}_q$ and $\widetilde{\rho}_q$ are the Fourier
transform of the tilt and roll variables (the two bending modes),
$\omega_0=1.76~\text{nm}^{-1}$ is the average intrinsic twist and
$\Delta q \equiv N a \omega_0/(2\pi)$. Equation~\eqref{eq:integral_lB}
has a similar $m$-dependence as \eqref{eq:integral}, but there are
some differences.  The bending in DNA is described by two components:
the (stiffer) tilt $\widetilde{\tau}_q$ and the (softer) roll
$\widetilde{\rho}_q$. In addition \eqref{eq:integral_lB} contains
$\omega_0$ (both explicitly and via $\Delta q$) since the DNA frame
rotates with the intrinsic twist and this needs to be taken into account
in the calculation of $l_B$. The expression \eqref{eq:integral_lB}
was derived under the assumption that the roll and tilt angles are
small compared to the intrinsic twist angle which is $\approx 34^\circ$
(an approximation referred to as intrinsic twist dominance\cite{skor21}).

Figure~\ref{fig:C-stiff}(b) shows a plot of the numerical estimate of 
\eqref{eq:integral_lB} for cgDNA and cgDNA+ (dashed lines). Length 
scale effects are here much weaker as compared to those observed in 
the torsional stiffness of Fig.~\ref{fig:C-stiff}(a). As for the twist,
also in bending the DNA appears to be softer at short length scales 
and asymptotically stiffer as $m \to \infty$, as also observed in other
coarse-grained DNA models \cite{skor21}. Again cgDNA+ is stiffer than 
cgDNA at high length scales. The weaker $m$-dependence of $l_B$ is due 
to the fact that the integrand $\Psi_q$ in \eqref{eq:integral_lB}
depends very weakly on $q$. As $\int \sin^2(my)/\sin^2(y) dy = m \pi$ a
weak $q$-dependence leads to a weak $m$-dependence of $l_B$. Bending 
fluctuations are dominated by the softer roll, i.e. $|\widetilde{\rho}_q| 
\gg |\widetilde{\tau}_q|$. As opposed to tilt, roll fluctuations very 
weakly depend on $q$, which is a typical feature of double stranded 
polymers \cite{sege22}. We have also computed the bending persistence 
length via decay of the tangent-tangent correlation functions
\begin{eqnarray}
    \langle \hat{t}_n \cdot \hat{t}_{n+m} \rangle = e^{-am/l_B(m)}.
\label{eq:lB_corr}
\end{eqnarray}
The data for $l_B$ vs. base pair distance $m$ thus obtained are shown
as solid lines in Fig.~\ref{fig:C-stiff}(b). There is a noticeable
difference in the extrapolated $l_B$ from \eqref{eq:integral_lB} and
\eqref{eq:lB_corr}, with the former underestimating the persistence
length. This discrepancy was observed previously\cite{skor21}
and the oscillatory behavior of the latter method stems from
the helicity of the traced contour. The extrapolated data for
cgDNA/cgDNA+, as well as the experimental value of $l_B$ are given in
Table~\ref{tab:persistence_lengths}.

One can notice that the asymptotic value found for the persistence 
length in case of cgDNA+ goes above the usually accepted value of 
$l_B \approx 50~nm$. This could be expected, as it was reported before 
that cgDNA predicts a persistence length that is slightly higher than 
experimental values \cite{mitc17}. Since we showed that cgDNA+ behaves 
stiffer at the long scale concerning the bending modes, this comes as no 
surprise. It is important to mention that although the bending persistence 
length of cgDNA+ is asymptotically deviating from the experimental value, 
the cgDNA+ curve is consistent with all-atom data\cite{skor21} when both 
are evaluated with Eq.~\eqref{eq:integral_lB}. This can be seen for the 
asymptotic value in Table~\ref{tab:persistence_lengths}.

\begin{table}[t]
    \centering
    \begin{tabular}{c|c|c|c|c}
        & cgDNA & cgDNA+ & All-Atom & Experiments\\
        \hline
        {$\cal{C}$} & $47$ / $47$ & $134$ / $133$ & $125$ & $110 \pm 10$\\
        {${l_B}$} & $53$ / $48$ & $72$ / $65$ & $61$ & $45 \pm 5$\\
    \end{tabular}
    \caption{
    Asymptotic values (data in $nm$) of the torsional stiffness ${\cal
    C}$ and of the bending persistence length $l_B$ as obtained from
    extrapolation of cgDNA/cgDNA+ simulations vs. all-atom\cite{skor21}
    and experimental data. The measured DNA bending persistence length
    is salt dependent and at physiological salt concentration $\sim
    150$~nM NaCl is about $l_B = 45 \pm 5$~nm \cite{bloo00}. The
    torsional stiffness was measured by several groups with slightly
    different results \cite{brya03,lipf10,gao21} and found to be
    salt independent. The value and error bar reported in the Table
    (${\cal C} = 110\pm 10$~nm) covers the range of literature
    values \cite{brya03,lipf10,gao21}.  For the MC simulations of
    cgDNA/cgDNA+ $100$ random sequences of length $500$~bp were used.
    For cgDNA/cgDNA+, the first value represents the stiffness found from
    the correlation functions (full lines in Fig.~\ref{fig:C-stiff}),
    while the second values is obtained from numerical integration (dotted
    lines in Fig.~\ref{fig:C-stiff}). These asymptotic stiffnesses were
    obtained by fitting the curves using Eq.~\eqref{eq: 1/m behaviour of
    C_m }, where the initial data points ($m < 20$) were excluded from
    the fit to probe the long length scales, as well as the final ones
    ($m > 300$) to exclude boundary effects. The all-atom values were
    taken from Ref.~\onlinecite{skor21}, which were obtained only by
    numerical integration.}
    \label{tab:persistence_lengths}
\end{table}

\section{Distal sites correlations}
\label{seq:correlations}

The non-local couplings induce correlations between coarse-grained
variables defined at DNA distal sites $n$ and $n+m$. These correlations
can be linked to the $q$-stiffness. We illustrate this for the
twist-twist correlation function, which via the discrete Fourier transform
\eqref{eq:energy_fourier} can be written as
\begin{equation}
    \left\langle  \Omega _n \Omega_{n+m} \right\rangle 
    = \frac{1}{N^2} \sum_q \left< |\widetilde{\Omega}_q |^2\right> 
    e^{-2\pi i m q/N} 
    = \frac{1}{N} \sum_q \frac{e^{-2\pi i m q/N}}{a\widetilde{\cal C}_q},
\label{eq:correl}
\end{equation}
where we have used the equipartition relation analogous to
\eqref{eq:covariance} to link twist fluctuations of the mode $q$
to $\widetilde{\cal C}_q$. The latter, as in \eqref{eq:en-twist},
is the twist stiffness of the mode $q$ obtained from integrating out
all other coarse-grained coordinates. We note that \eqref{eq:correl}
is different from the twist correlation function discussed above and
defined as the average of the cosine of the twist angle (see caption of
Fig.~\ref{fig:C-stiff}).  The latter does not vanish even in absence
of non-local couplings, while in that case $\widetilde{\cal C}_q$
is $q$-independent and the last term in \eqref{eq:correl} becomes
a sum over $q$ of exponential phases $\exp(-2\pi i m q/N)$ which
vanishes for any $m\neq 0$.  The asymptotic decay of \eqref{eq:correl}
is governed by the leading pole of $\widetilde{\cal C}_q$, i.e. the
solution of $\widetilde{\cal C}_q=0$ with the smallest imaginary part
\cite{sege23}. Such equation cannot have a real solution $q=q^*$ as this
would imply an unstable mode. The most generic pole is of the form
\begin{eqnarray}
    \frac{2\pi q_E}{N} &=& \phi \pm \frac{i}{\xi_E},
\label{eq:qE}
\end{eqnarray}
which leads to the following asymptotic, damped oscillatory decay of
the normalized correlator \cite{sege23}
\begin{eqnarray}
    \frac{\left\langle  \Omega _n \Omega_{n+m} \right\rangle }
    {\left\langle \Omega_n^2 \right\rangle} &\stackrel{m \gg 1}{\sim}&
    \cos(m\phi +\phi_0) \, e^{-m/\xi_E},
\label{eq:twist-twist_correl}
\end{eqnarray}
with $\phi_0$ a phase factor. The real part of the leading pole
\eqref{eq:qE} gives the oscillation frequency, while the imaginary part
is the inverse decay length, hence  the pole with the smallest imaginary
part corresponds to the slowest decaying mode.

\begin{figure}[t]
    \centering
    \includegraphics[width=\linewidth]{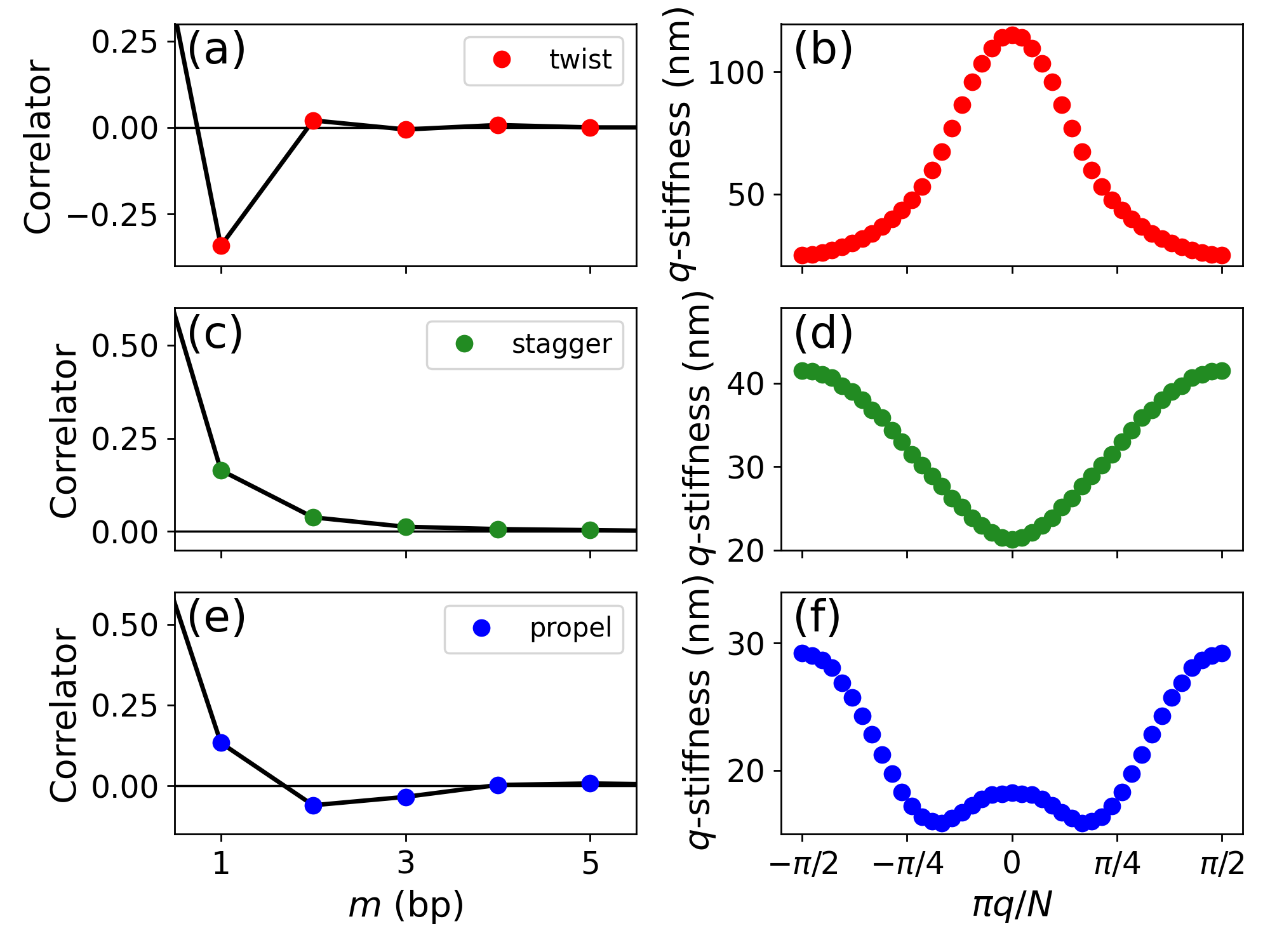}
\caption{(a,c,e) Normalized real-space correlation function for various
degrees of freedom as obtained from cgDNA+ for a 44-mer. (b,d,f)
Corresponding $q$-stiffness spectra. The $q$-stiffnesses are obtained
for each variable separately (see Tab.~\ref{tab:seqs}), e.g. they are
Schur-complements of the $12\times12$ matrix in Eq. \eqref{eq:Mq}
by projecting the 12-dimensional configuration space onto a single
dimension.}
\label{fig:correlation_functions} 
\end{figure}

Figure~\ref{fig:correlation_functions} shows (a) the normalized
twist-twist correlator \eqref{eq:twist-twist_correl} and (b) the
corresponding twist stiffness $\widetilde{\cal C}_q$ for cgDNA+. As
a comparison, we also show in Fig.~\ref{fig:correlation_functions}
the normalized correlator and associated stiffness for two other
intra-bp variables: stagger (c,d) and propeller twist (e,f). The
stagger is the relative displacement of the two bases forming a base
pair along the double helix axis. The propeller twist is the relative
rotation angle of the two bases in a base pair. In all three cases
shown, the correlator decay reflects the stiffness behavior. In
first approximation, the poles real part corresponds to the minimum
of the stiffness. In the twist, stagger and propeller twist cases the
minimum would give $\phi=\pi$, $\phi=0$ and $\phi \approx \pi/3$. The
first case corresponds to a correlator decay modulated by a factor
$\cos(m\pi)=(-1)^m$, i.e. an alternated sign. The second case ($\phi=0$)
leads to a monotonous decay, while the third case ($0 < \phi < \pi$)
would correspond to an oscillating decay. The correlators plotted in
Fig.~\ref{fig:correlation_functions}(a,c,e) reproduce indeed these three
distinct behaviors.

We note that correlators rapidly decay and virtually vanish at the
distance of $m=4$~bp. Such correlations potentially generate effective
interactions at distal sites which are linked to allosteric effects
\cite{sege23}. Experiments have shown that proteins binding at distal DNA
sites interact with each other via the linker DNA \cite{kim13,rose21}.
The characteristic decay length observed in experiments is about $\sim
15$~bp, which is about one order of magnitude larger than the decay shown
in Fig.~\ref{fig:correlation_functions}(a,c,e).  This discrepancy suggests
that distal correlations in DNA are stronger than all-atom simulations
(and cgDNA+) currently predict \cite{sege23}.

\section{Discussion}
\label{sec:discussion}

In this paper we have analyzed the properties of a class of coarse-grained
DNA models known as cgDNA\cite{petk14} and the most recent improved
version cgDNA+\cite{pate19,shar23}. In these models each base is treated
as a rigid object. Translations and rotations between two bases forming
a base pair are parametrized by six intra-bp coordinates, while relative
orientations between consecutive base pairs are defined by another set
of six inter-bp coordinates, see Fig.~\ref{fig:snapshot-cgDNA}.  In this
way each site in cgDNA is parameterized by $12$ coordinates, for cgDNA+
phosphate interactions are added which brings it to $24$ coordinates
per site\cite{pate19}. We compared the two models by integrating out
the phospate contribution of cgDNA+ focusing on the calculation of the
$12 \times 12$ stiffnesses matrix in Fourier space $\widetilde{M}_q$.
The $q$-dependence of the elements of the stiffness matrix reflects the
existence of distal couplings, e.g. coupling between non-proximal sites
\cite{skor21,sege22,fosa23}.

Our analysis shows that cgDNA+ is a significant improvement of cgDNA
in several aspects. As illustrated in Fig.~\ref{fig:q-stiff}(a-c),
the tilt- and twist coordinates of cgDNA+ show a high level of overlap
with all-atom data over the whole $q$-stifness range. This means
that cgDNA+ accurately encodes both the short and long length-scale
behavior of the all-atom simulations. The cgDNA model on the other
hand, only captures the short length scale of tilt and twist. The
improvement of cgDNA+ in twist behavior is especially outspoken, with the
torsional stiffness data showing a very strong length-scale dependence
(Fig.~\ref{fig:C-stiff}(a)), consistent with experiments indicating a
torsionally softer DNA at short distances \cite{skor21}. Moreover, its
asymptotic value for the long length scale is close to the experimental
value obtained from the measurement of Kbp long DNA\cite{lipf14}
(see Table~\ref{tab:persistence_lengths}).  Conversely, the bending
persistence length of cgDNA+ more significantly overestimates the
consensus experimental value as shown from the data reported in
Table~\ref{tab:persistence_lengths}. The excellent overlap between
all-atom and cgDNA+ $q$-stiffnesses suggests that the problem with the
bending persistence length is possibly due to the current force fields
used in all-atom models\cite{lieb23}, rather than a problem with the
parametrization of cgDNA+.

Finally, as a methodology, the analysis of $q$-stiffnesses can be very
useful in the development of coarse-grained DNA models. A good match
of stiffnesses at all $q$-values with respect to some reference data
is an indication of a correct parametrization accounting both for the
short- and long distance behavior, as illustrated in the case of the
torsional stiffness. Problematic behavior of some degrees of freedom can
be spotted more easily from the analysis of $q$-stiffnesses.  This is
illustrated in the case of the inter-bp rotation coordinates tilt
and twist of cgDNA in Fig.~\ref{fig:q-stiff}(a). From a computational
point-of-view, it is important to note that the $q$-stiffness analysis
can be performed reliably on systems of rather small lengths, see Appendix
\ref{appendix:finite size}, since it converges rapidly to its asymptotic
values over the whole $q$-range.

\acknowledgments{Discussions with E. Skoruppa are gratefully acknowledged.
M. S. acknowledges financial support from Fonds Wetenschappelijk Onderzoek
(FWO 11O2323N).  }


\appendix

\section{All-atom and cgDNA(+) simulations}
\label{appendix:all_atom}

MD all-atom simulations were performed on the two $44$~bp sequences given
in Table~\ref{tab:seqs} and as described in Ref.~\onlinecite{sege23}.
Briefly, Gromacs v.2020.4 \cite{abra15} and the Amberff99 parmbsc1 force
field \cite{ivan16} were used. The TIP-3P water model \cite{jorg83}
was used while non-bonded interactions were cutoff at $1.0$~nm and PME
Mesh Ewald interactions for electrostatic. The molecules were placed in
a dodecahedral box with $2.0$~nm spacing on the sides using periodic
boundary conditions and a neutralized solution of $150$~mM NaCl.
The energy was minimized with tolerance of $1000$~kJ/mol. A run of
$100$~ps at $300$~K using a velocity rescaling thermostat \cite{buss07}
was followed by $100$~ps at the same temperature using a Parrinello-Rahman
barostat \cite{parr81} with $1.0$~bar pressure. Production runs were
performed using the latter barostat and run for $100$~ns using a $2$~fs
time step in a leapfrog integrator. The trajectories were analyzed using
the software Curves+ \cite{lave09}, which extracts the twelve rigid base
coordinates from atomic configurations. To minimize endpoint effects two
base pairs on each end of the molecule were excluded from the analysis.

\begin{table}[t]
    \centering
    \begin{tabular}{c}
    Sequences \\
    \hline
     \footnotesize{{CGCTCAAGGGCGAGAATTGGACCTGGCTTACGTCTTAGTACGTA}$^*$}\\
    \footnotesize{GTTAAGTGCCGAACTAGATCTGACCTAACGGTAAGAGAGTTTCA}\\
    \end{tabular}
    \caption{The two $44$~bp sequences used in the all-atom MD
    simulations and in the cgDNA/cgDNA+ models to produce the data 
    shown in Figures~\ref{fig:q-stiff} and \ref{fig:q-stiff-imag}.
    The sequence indicated with $*$ was used along with the cgDNA+
    model to generate the data in Fig. \ref{fig:correlation_functions}.}
    \label{tab:seqs}
\end{table}

The data of cgDNA/cgDNA+ models were obtained from Monte Carlo
simulations. For cgDNA we used the 2018 2.0 version using the parameter
set cgDNAps4. For cgDNA+ we used the parameter set DNA PS2. To obtain
the $q$-stiffnesses of Fig.~\ref{fig:q-stiff}, \ref{fig:q-stiff-imag}
we sampled $5 \times 10^5$ configurations for each of the two $44$-mer
DNA sequences of Table~\ref{tab:seqs}.  For Fig.~\ref{fig:C-stiff},
100 different sequences of length 500 bp were used ($10^3$ samples per
sequence), in order to capture to long length scale directly from the
correlation functions. As in the all-atom case, the two outer base pairs
from each side of the sequences were removed from the analysis of the
stiffnesses to reduce end-effects \cite{skor21}.  The cgDNA/cgDNA+-output
for rotational coordinates uses a Cayley vector representation, deviating
from the Euler vector representation 
which was used for the analysis of all-atom MD simulations. Given a
Cayley vector $\vec{\theta}$, it can be converted to an Euler vector
$\vec{\omega}$ using Eq. \eqref{eq:Cay_To_Euler}
\begin{equation}
    \vec{\omega}=2\arctan \left(\frac{|\vec{\theta}|}{2}\right)
    ` \frac{\vec{\theta}}{|\vec{\theta}|}.
    \label{eq:Cay_To_Euler}
\end{equation}
In order to compare $q$-stiffnesses from cgDNA/cgDNA+ with those obtained
from all-atom simulations, rotational coordinates from cgDNA/cgDNA+
were first converted into Euler vectors prior to the calculation of
momentum-space stiffnesses.

\begin{figure}[t]
    \centering
    \includegraphics[width=\linewidth]{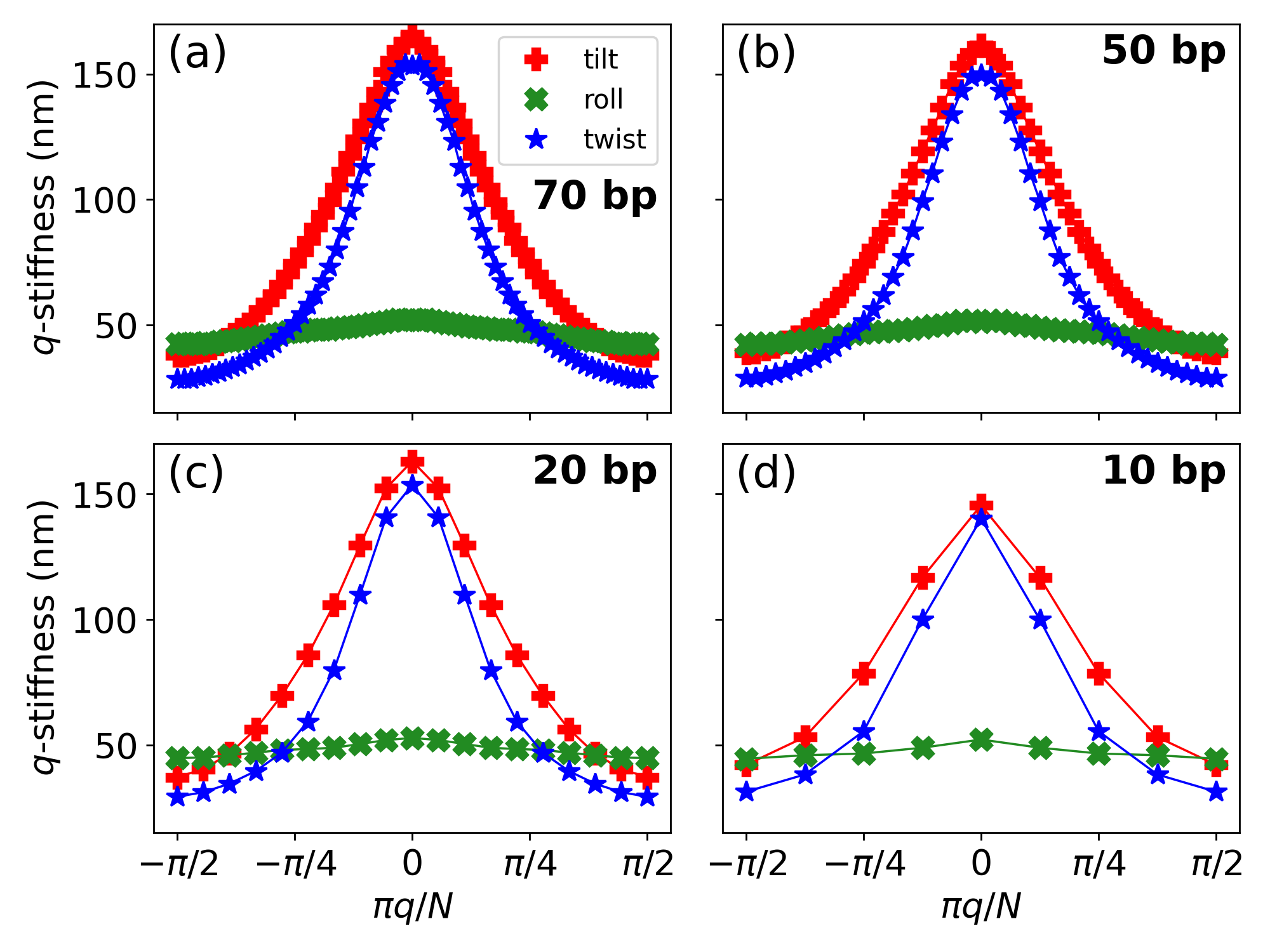}
\caption{$q$-stiffnesses for tilt, roll and twist for various DNA
lengths.  The $q$-stiffnesses are obtained by considering the three
dimensional configuration space spanned by tilt, roll and twist degrees
of freedom. In the tilt-roll-twist space, the stiffness matrix becomes
a $3\times 3$-matrix which forms a Schur-complement of the $12\times12$
matrix in Eq.~\eqref{eq:Mq}.}
\label{fig:finite size} 
\end{figure}

\section{Finite size effects}
\label{appendix:finite size}

One of the remarkable features we noticed of the $q$-stiffnesses
is their rapid convergence to asymtptotic values already for rather
short molecules. We illustrate this behavior for tilt, roll and twist
stiffnesses in Fig.~\ref{fig:finite size}, which shows cgDNA+ calculations
for molecules of different lengths ($10$, $20$, $50$ and $70$ bp). Plotted
are the diagonal elements of the $3 \times 3$ stiffness matrix describing
tilt, roll and twist deformations where all other degrees of freedom
are integrated out.  Already the $10$~bp dataset shows stiffnesses which
develop peaks at $q=0$ which are remarkably close to the long molecule
limit values. This suggests that from the analysis of the $q$-stiffnesses
one can reliably extract information concerning the elastic behavior,
and their length-scale dependence, of very long molecules.


%

\end{document}